\documentclass[twocolumn]{aastex631}

\received{ }
\revised{ }
\accepted{}
\submitjournal{ApJ}

\shorttitle{Temporal evolution of non-neutralized electric currents}
\shortauthors{Kontogiannis \& Georgoulis}

\graphicspath{{./}{figures/}}

\begin{document}

\title{The temporal evolution of non-neutralized electric currents and the complexity of solar active regions}

\correspondingauthor{Ioannis Kontogiannis}
\email{ikontogiannis@aip.de}

\author[0000-0002-3694-4527]{Ioannis Kontogiannis}
\affiliation{Leibniz-Institut f\"{u}r Astrophysik Potsdam (AIP)
An der Sternwarte 16, 14482 Potsdam, Germany}

\author[0000-0001-6913-1330]{Manolis K. Georgoulis}
\affiliation{Space Exploration Sector, Johns Hopkins Applied Physics Laboratory, Laurel, MD 20723, USA}
\affiliation{Research Center for Astronomy and Applied Mathematics of the Academy of Athens, 11527 Athens, Greece}

\begin{abstract}
We study the evolution of electric currents during the emergence of magnetic flux in the solar photosphere and the differences exhibited between solar active regions of different Hale complexity classes. A sample of 59 active regions was analyzed using a method based on image segmentation and error analysis to determine the total amount of non-neutralized electric current along their magnetic polarity inversion lines. The time series of the total unsigned non-neutralized electric current, $I_{NN,tot}$, exhibit intricate structure in the form of distinct peaks and valleys. This information is largely missing in the respective time series of the total unsigned vertical electric current $I_{z}$. Active regions with $\delta$-spots stand out, exhibiting 1.9 times higher flux emergence rate and 2.6 times higher $I_{NN,tot}$ increase. The median value of their peak $I_{NN,tot}$ is equal to $3.6\cdot10^{12}$\,A, which is more than three times higher than that of the other regions of the sample. An automated detection algorithm was also developed to pinpoint the injection events of non-neutralized electric current. The injection rates and duration of these events were higher with increasing complexity of active regions, with regions containing $\delta$-spots exhibiting the strongest and longest events. These events do not necessarily coincide with increasing magnetic flux, although they exhibit moderate correlation. We conclude that net electric currents are injected during flux emergence, but are also shaped drastically by the incurred photospheric evolution, as active regions grow and evolve. 

\end{abstract}


\section{Introduction} 
\label{sec:intro}

Active regions form when buoyant magnetic flux tubes rise from the interior of the Sun \citep{2015LRSP...12....1V}, and appear in the photosphere, forming pores, sunspots and/or groups of sunspots \citep{1985SoPh..100..397Z,1987ARA&A..25...83Z} of varying size and complexity. Their white light morphology and distribution of magnetic polarities can be used to categorize them in terms of magnetic complexity \citep{1919ApJ....49..153H}. Active regions 
can be as simple as isolated sunspots surrounded by plage regions ($\alpha$-type), bipolar or multipolar sunspot groups consisting of well-, or not so well-, separated polarities ($\beta$- and $\beta\gamma$-type), and in some cases they include opposite-polarity umbrae sharing the same penumbra \citep[$\delta$-type regions,][]{1965AN....288..177K}. Different complexity categories can also reflect different evolutionary states, as active regions emerge and evolve \citep{1990SoPh..125..251M}. Although most regions are relatively simple and classified as $\beta$ \citep[bipolar;][]{2016ApJ...820L..11J}, the most complex active regions 
($\delta$-spots) exhibit sheared polarity inversion lines (PILs) characterized by exceptionally strong electric currents and are the ones most often associated with strong flares and coronal mass ejections (CMEs) \citep{2000ApJ...540..583S,2019LRSP...16....3T}.

The existence of strong electric currents owing to complex magnetic fields was noticed already several decades ago \citep{1965SvA.....9..171S,1967SoPh....1..220A}. Although concerns were raised against the realism and feasibility of calculation of these currents
\citep{parker96,wilkinson92}, when appropriate spectropolarimetric observations became commonplace, strong indications occurred that the inferred currents from Ampere's law are reasonably realistic and reflect true physical characteristics \citep{Leka96,1998A&A...331..383S, 2000ApJ...532..616W,geo_titov_mikic12}. These studies further showed that some active region electric currents exhibited non-neutrality, meaning that their 
photospheric magnetic footprints deviated from the archetypal 
isolated shielded flux tubes predicted by theory \citep{parker79} and discussed in studies 
of isolated magnetic elements \citep{2009ApJ...706L.114V}. A theoretical framework as well as merits and caveats of the study of electric currents in solar active regions can be found in \citet{2018GMS...235..371G}. In the nominal case of isolated flux tubes embedded in a field-free plasma, a volume (direct) electric current is shielded by an equal and oppositely directed (return) sheath current. This symmetry possibly  breaks in cases of very close proximity of opposite-polarity footprints such as, prominently, intense magnetic polarity inversion lines. In such exceptional cases, electric currents are non-neutralized and raw currents are injected in the solar atmosphere, supplying the otherwise non-neutralized currents of the overwhelmingly space-filling coronal magnetic fields \citep{longcope_welsch00}.

Recent studies have showcased the importance of the electric currents in the context of the pre-eruption evolution of solar active regions \citep{2019RSPTA.37780094G}, as well as the three-dimensional standard flare model \citep{2012A&A...543A.110A,2013A&A...549A..66A,2013A&A...555A..77J,janvier14}. Overall, flaring active regions exhibit at least an order of magnitude higher non-neutralized electric currents than flare-quiet ones \citep{kontogiannis17,kontogiannis18} and the same quantity is highly correlated with the kinematic characteristics of CMEs \citep{kontogiannis19}. Regions prone to eruptions are reported to exhibit a higher degree of non-neutralization \citep{2017ApJ...846L...6L,2019MNRAS.486.4936V,2020ApJ...893..123A,2024ApJ...961..148L}. Localized, intense changes of the vertical electric current density have also been associated with pre-flaring activity and free energy build-up \citep{2020SoPh..295...29M, suraj2023}. Flux-ropes with non-neutralized electric currents at their footpoints may well be kink-unstable \citep{2022ApJ...939..114T}, which seems to be corroborated by at least some  observed flaring active regions \citep{2020ApJ...895...18B,2023ApJ...943...80W}. 

But what is the origin of non-neutralized electric currents in solar active regions?  Simulations show that the development of non-neutralized electric currents is a consequence of flux emergence and the subsequent photospheric motions, which lead to the formation of PIL and strong magnetic shear \citep{torok14,dalmasse15}. The coexistence of strong electric currents and PILs is substantiated by observational studies \citep[see e.g.,][]{kontogiannis19,2022ApJ...926...56K}. On the other hand, \citet{geo_titov_mikic12} propose that the close interaction between opposite magnetic polarities leads to the breaking of cylindrical symmetry and the development of shear and strong electric currents through the Lorentz force. In the same study, the coherence found in the electric current carried by the magnetic polarities points to a sub-photospheric origin of electric currents. What is, then, the evolution of the electric current in active regions and what conclusions can be drawn from its scrutiny?  This question is relevant to how some active regions eventually become so complex. Although several studies have focused on the evolution of the magnetic flux during flux emergence \citep{2011PASJ...63.1047O, 2017ApJ...842....3N, 2019MNRAS.484.4393K}, no study so far has been dedicated to the evolution of non-neutralized electric currents in a sizeable sample of emerging active regions. Smaller samples of regions/events have been scrutinized using the degree of current neutralization \citep[e.g.,][]{2017ApJ...846L...6L,2020ApJ...893..123A,2020ApJ...900...38H,2023ApJ...943...80W,2024ApJ...961..148L}, whose calculation is based on assumptions regarding direct and return currents. 

In the present study we examine a statistically significant sample of emerging active regions of different magnetic complexity (in terms of Hale class), following a methodology (Section~\ref{sec:obs_anal}) that calculates the non-neutralized electric current using morphological image processing and taking into account the systematic and numerical errors as in \citet[][]{geo_titov_mikic12}.  
In Section~\ref{sec:results}, we discuss a few 
 case studies of specific active regions
emphasizing the evolution of electric current in different complexity classes. We then present the characteristics of the non-neutralized electric currents time series in comparison to the time series of the magnetic flux and as a function of Hale class. Additionally, we analyze the time series in terms of injection events and their properties. In Section~\ref{sec:conclusions}, we summarize and discuss the results in the context of the origin of electric currents in active regions and the formation of $\delta$-spot regions.

\startlongtable
\begin{deluxetable*}{ccccccccc}
\tabletypesize{\scriptsize}
\tablecaption{The sample of emerging active regions used in this study. The table contains information regarding their National Oceanic and Atmospheric Administration (NOAA) identifier, the start and end date of each HARP data set, the start and end heliographic longitude at which each region was observed, the resulting number of processed magnetograms (the number of points in the time series), the peak Hale class each region reached over the observation interval, and the number of flares associated with each region (see text). }

\tablenum{1}

\tablehead{\colhead{Number} & \colhead{NOAA} & \colhead{Start Date} & \colhead{End Date} & \colhead{Start long.} & \colhead{End long.} & \colhead{Number of} & \colhead{Hale class} & \colhead{Flare class} \\ 
\colhead{} & \colhead{} & \colhead{} & \colhead{} & \colhead{(Degrees)} & \colhead{(Degrees)} & \colhead{magnetograms} & \colhead{} & \colhead{}} 

\startdata
       1 &  11440  &  19-Mar-2012 00:46  &  26-Mar-2012 19:10  &  -22.87  &  93.00  & 853 & $\beta\gamma\delta$ & 4\,C\\
       2 &  11267  &   4-Aug-2011 10:10  &  12-Aug-2011 11:22  &  -46.62  &  58.76  & 963 & $\beta\gamma\delta$ & 1\,B, 4\,C\\
       3 &  11465  &  19-Apr-2012 14:46  &  30-Apr-2012 13:10  &  -57.92  &  90.97  & 1300 & $\beta\gamma\delta$ & 3\,B, 16\,C\\
       4 &  11640  &  29-Dec-2012 12:10  &   8-Jan-2013 10:46  &  -40.28  &  87.82  & 1191 & $\beta\gamma\delta$ & 7\,B, 7\,C\\
       5 &  11726  &  19-Apr-2013 04:22  &  27-Apr-2013 09:10  &  -19.41  &  88.68  & 924 & $\beta\gamma\delta$ & 11\,B, 56\,C, 1\,M\\
       6 &  11158  &  10-Feb-2011 21:58  &  21-Feb-2011 07:34  &  -40.26  &  92.56  & 1247 & $\beta\gamma\delta$ & 1\,B, 56\,C, 5\,M, 1\,X\\
       7 &  11750  &  15-May-2013 00:58  &  21-May-2013 18:58  &  0.4265  &  90.19  & 808 & $\beta\gamma\delta$ & 6\,C\\
       8 &  12673  &  28-Aug-2017 08:58  &  10-Sep-2017 11:10  &  -81.77  &  89.12  & 1431 & $\beta\gamma\delta$ & 1\,B, 54\,C, 26\,M, 4\,X\\
       9 &  11515  &  26-Jun-2012 03:58  &  10-Jul-2012 03:10  &  -81.46  &  88.61  & 1668 & $\beta\gamma\delta$ & 2\,B, 73\,C, 30\,M, 1\,X\\
      10 &  11884  &  26-Oct-2013 06:58  &   8-Nov-2013 10:46  &  -83.82  &  88.84  & 1571 & $\beta\gamma\delta$ & 1\,B, 15\,C, 4\,M\\
      11 &  12257  &   4-Jan-2015 03:46  &  14-Jan-2015 14:46  &  -50.51  &  89.60  & 1252 & $\beta\gamma\delta$ & 23\,C, 3\,M\\
      12 &  11561  &  30-Aug-2012 00:22  &   5-Sep-2012 20:10  &  -29.26  &  62.02  & 727 & $\beta\gamma\delta$ & \\
      13 &  11645  &   2-Jan-2013 19:34  &   8-Jan-2013 23:10  &  -12.42  &  75.48  & 736 & $\beta\gamma\delta$ & 2\,C\\
      14 &  11431  &   4-Mar-2012 11:10  &  10-Mar-2012 09:22  &  14.617  &  94.20  & 681 & $\beta\gamma\delta$ & \\
      15 &  11273  &  16-Aug-2011 13:10  &  21-Aug-2011 16:34  &  -18.69  &  52.27  & 618 & $\beta\gamma\delta$ & \\
      16 &  11179  &  21-Mar-2011 09:58  &  26-Mar-2011 22:22  &  -16.62  &  61.42  & 606 & $\beta\gamma$ & \\
      17 &  11327  &  18-Oct-2011 23:58  &  28-Oct-2011 16:34  &  -38.71  &  87.77  & 1107 & $\beta\gamma$ & \\
      18 &  11813  &   6-Aug-2013 14:10  &  12-Aug-2013 11:34  &  -13.30  &  69.01  & 699 & $\beta\gamma$ & 1\,B\\
      19 &  11776  &  18-Jun-2013 11:34  &  26-Jun-2013 00:58  &  -11.47  &  90.45  & 907 & $\beta\gamma$ & 5\,B, 5\,C\\
      20 &  11824  &  17-Aug-2013 06:10  &  22-Aug-2013 07:10  &  2.1014  &  75.38  & 597 & $\beta\gamma$ & 1\,C\\
      21 &  11922  &  10-Dec-2013 00:22  &  15-Dec-2013 19:22  &  4.4104  &  89.78  & 696 & $\beta\gamma$ & \\
      22 &  12353  &  19-May-2015 11:10  &  27-May-2015 22:10  &  -46.09  &  74.15  & 1008 & $\beta\gamma$ & 1\,B, 3\,C\\
      23 &  11416  &   8-Feb-2012 13:58  &  18-Feb-2012 10:22  &  -41.35  &  92.18  & 1060 & $\beta\gamma$ & 10\,B, 1\,C\\
      24 &  11682  &  23-Feb-2013 19:58  &   6-Mar-2013 06:34  &  -37.24  &  00.00  & 1135 & $\beta\gamma$ & 5\,B\\
      25 &  11781  &  27-Jun-2013 20:10  &   5-Jul-2013 01:46  &  -11.43  &  82.70  & 859 & $\beta\gamma$ & 4\,B, 1\,C\\
      26 &  11946  &   4-Jan-2014 04:58  &  14-Jan-2014 05:34  &  -47.10  &  89.12  & 1196 & $\beta\gamma$ & 3\,C, 1\,M\\
      27 &  12003  &   9-Mar-2014 15:34  &  16-Mar-2014 22:10  &  -7.307  &  89.50  & 765 & $\beta\gamma$ & 9\,C\\
      28 &  12271  &  24-Jan-2015 06:22  &   2-Feb-2015 14:22  &  -32.58  &  87.96  & 1087 & $\beta\gamma$ & 1\,B, 4\,C\\
      29 &  12339  &   4-May-2015 16:34  &  18-May-2015 17:58  &  -84.50  &  89.56  & 1636 & $\beta\gamma$ & 3\,B, 56\,C, 3\,M, 1\,X\\
      30 &  12543  &   7-May-2016 16:58  &  16-May-2016 17:58  &  -30.72  &  90.04  & 1043 & $\beta\gamma$ & 7\,B, 5\,C\\
      31 &  11079  &   8-Jun-2010 03:34  &  13-Jun-2010 13:46  &  12.102  &  89.36  & 589 & $\beta\gamma$ & 1\,B, 1\,M\\
      32 &  11326  &  20-Oct-2011 03:34  &  24-Oct-2011 15:46  &  23.325  &  86.04  & 534 & $\beta\gamma$ & \\
      33 &  11385  &  22-Dec-2011 02:58  &  28-Dec-2011 22:22  &  -22.34  &  63.62  & 811 & $\beta\gamma$ & \\
      34 &  11472  &  29-Apr-2012 04:10  &   8-May-2012 00:34  &  -51.84  &  60.73  & 915 & $\beta\gamma$ & \\
      35 &  11807  &  28-Jul-2013 10:22  &   3-Aug-2013 12:58  &  -2.951  &  81.79  & 729 & $\beta\gamma$ & 1\,B\\
      36 &  11081  &  11-Jun-2010 06:22  &  14-Jun-2010 23:22  &  35.170  &  90.15  & 439 & $\beta\gamma$ & 6\,B, 7\,C, 1\,M \\
      37 &  11148  &  17-Jan-2011 01:58  &  21-Jan-2011 09:46  &  21.677  &  79.32  & 509 & $\beta\gamma$ & \\
      38 &  11116  &  16-Oct-2010 19:22  &  19-Oct-2010 02:58  &  -2.952  &  30.11  & 259 & $\beta\gamma$ & \\
      39 &  11510  &  18-Jun-2012 19:58  &  25-Jun-2012 04:22  &  -18.74  &  71.90  & 713 & $\beta\gamma$ & \\
      40 &  11141  &  30-Dec-2010 21:34  &   6-Jan-2011 01:34  &  -1.415  &  81.40  & 736 & $\beta$ & 10\,B, 1\,C\\
      41 &  11072  &  20-May-2010 16:22  &  29-May-2010 12:34  &  -35.76  &  90.09  & 1050 & $\beta$ & 4\,B\\
      42 &  11076  &  31-May-2010 04:10  &   7-Jun-2010 21:34  &  -16.68  &  89.09  & 925 & $\beta$ & 1\,B\\
      43 &  11096  &   3-Aug-2010 12:46  &  14-Aug-2010 02:10  &  -63.10  &  74.95  & 1189 & $\beta$ & 2\,B\\
      44 &  11132  &   3-Dec-2010 23:10  &  11-Dec-2010 10:22  &  -14.14  &  89.78  & 874 & $\beta$ & 4\,B\\
      45 &  11143  &   6-Jan-2011 00:46  &  13-Jan-2011 12:46  &  -43.12  &  54.76  & 901 & $\beta$ & \\
      46 &  11449  &  28-Mar-2012 09:10  &   3-Apr-2012 03:34  &  -3.404  &  78.27  & 680 & $\beta$ & \\
      47 &  11551  &  20-Aug-2012 03:46  &  25-Aug-2012 16:58  &  -12.12  &  62.68  & 656 & $\beta$ & \\
      48 &  11843  &  17-Sep-2013 07:46  &  23-Sep-2013 01:58  &  -14.49  &  65.78  & 637 & $\beta$ & \\
      49 &  11855  &  29-Sep-2013 18:10  &   7-Oct-2013 12:58  &  -32.89  &  78.70  & 896 & $\beta$ & \\
      50 &  12119  &  18-Jul-2014 09:34  &  26-Jul-2014 04:46  &  -23.16  &  84.18  & 867 & $\beta$ & 1\,B\\
      51 &  11152  &   2-Feb-2011 15:22  &   8-Feb-2011 08:46  &  -29.06  &  46.96  & 688 & $\beta$ & 8\,B\\
      52 &  11630  &   6-Dec-2012 17:10  &  15-Dec-2012 08:22  &  -47.67  &  71.92  & 1028 & $\beta$ & 11\,B, 2\,C\\
      53 &  11211  &   8-May-2011 15:10  &  11-May-2011 15:22  &  3.4807  &  45.80  & 356 & $\beta$ & \\
      54 &  11446  &  22-Mar-2012 16:58  &  27-Mar-2012 19:58  &  -17.52  &  52.96  & 570 & $\beta$ & 2\,B\\
      55 &  11105  &   2-Sep-2010 02:10  &   8-Sep-2010 12:46  &  -0.349  &  92.62  & 772 & $\beta$ & 13\,B, 2\,C\\
      56 &  11130  &  27-Nov-2010 15:10  &   5-Dec-2010 12:22  &  -19.48  &  89.99  & 946 & $\beta$ & 26\,B, 1\,C\\
      57 &  11242  &  28-Jun-2011 02:10  &   5-Jul-2011 07:58  &  -9.091  &  90.76  & 860 & $\beta$ & 7\,B,\\
      58 &  11300  &  17-Sep-2011 00:34  &  22-Sep-2011 10:58  &  17.306  &  93.13  & 594 & $\beta$ & \\
      59 &  11460  &  15-Apr-2012 11:34  &  26-Apr-2012 13:22  &  -61.20  &  88.76  & 1290 & $\beta$ & 2\,B, 5\,C\\
\enddata
\label{table:1}
\end{deluxetable*}

\section{Data and analysis}\label{sec:obs_anal}

\subsection{Data}

The Helioseismic and Magnetic Imager \citep[HMI;][]{hmischerrer,hmischou} onboard the Solar Dynamics Observatory \citep[SDO;][]{sdo} provides uninterrupted, full-disk, spectropolarimetric observations of the photosphere in the Fe\,I\,6173\,\AA\ spectral line since 2010. The spectropolarimetric observations are inverted to provide the magnetic field vector at one level in the photosphere. This study utilizes the HMI Active Region Patches \citep[HARP;][]{2014SoPh..289.3483H}, which are cutouts that track regions of interest (high-polarization regions such as active regions and plages) and contain, among others, maps of the three components of the magnetic field. The Cylindrical Equal Area (CEA) version of the definitive data is used, that is, cutouts for which the 180\degr - ambiguity has been resolved and the three components of the magnetic field vector $B_{r}$, $B_{p}$ and $B_{t}$ (equivalent to the three Cartesian components $B_{z}$, $B_{x}$ and $B_{y}$) have been deprojected and remapped to the solar disk center.

\begin{figure*}[htp!]
\centering
\includegraphics[width=18cm]{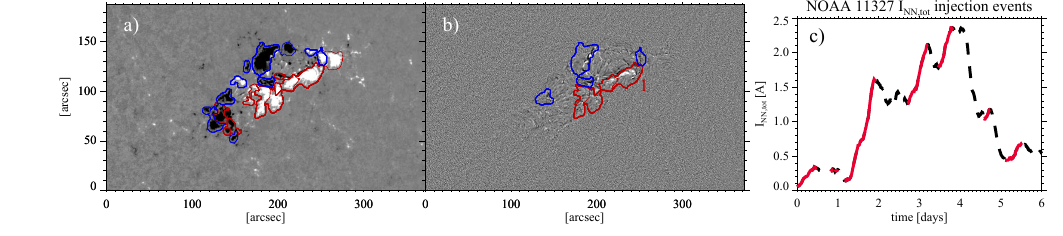}
\centering
\caption{Illustration of the partitioning process, the calculation of the non-neutralized electric current, and the automated injection event detection method. a) Map of the $B_{r}$ component for NOAA 11158, at 01:58 UT on 13 February 2011. The magnetic field values are clipped between $\pm$500\,G to enhance visibility. The red (blue) contours indicate magnetic partitions with positive (negative) total electric current. b)  Map of the vertical electric current density $J_z$, clipped between $\pm7\cdot10^{-6}$\,A\,cm$^{-2}$ to enhance visibility.
The red (blue) contours indicate magnetic partitions with  significant positive (negative) total electric current as per our selection criteria. The contour indicated with ``1'' is the partition with the maximum (in absolute value) non-neutralized electric current. c) Injection events (solid red) detected for the $I_{NN,tot}$ time series of NOAA\,11327, up to t = 6\,d. }
\label{fig:method}
\end{figure*}

A sample of 59 emerging active regions was studied, consisting of 20 $\beta$-, 24 $\beta\gamma$- and 15 $\beta\gamma\delta$-class regions (Table~\ref{table:1}). These Hale classes were the peak complexity classes reached by the active regions while they traversed the Earth-facing solar disk. The magnetic complexity information was derived from the Space Weather Prediction Center (SWPC) of the National Oceanic and Atmospheric Administration (NOAA)\footnote{\url{https://www.swpc.noaa.gov/}}. Information regarding the flaring activity of the regions was retrieved from the standardized multivariate data set SWAN-SF \citep{2020NatSD...7..227A}. 

\subsection{Derivation of non-neutralized electric currents}

The method to derive the net electric currents is the one developed by \citet{geo_titov_mikic12} and is illustrated schematically in Figure~\ref{fig:method}a and \ref{fig:method}b. The map of the vertical component of the magnetic field, $B_{r}$, is partitioned using a flux tessellation method \citep{barnes05}, with thresholds for $B_{r}$, minimum magnetic flux per partition and minimum partition size equal to 100\,G, 5$\times10^{19}$\,Mx and 40 pixels ($5.3\,Mm^2$), correspondingly. For each partition, the total current is calculated via a surface integration of the electric current density, which is itself calculated using the differential form of Ampere's law, rather than directly the
integral form  of Ampere's law used by \citet{geo_titov_mikic12}. The reason for this was to speed up the calculations. In \citet{kontogiannis17} it was shown that the difference between the two estimates were largely insignificant, in comparison to the accuracy of the magnetic field measurements. Then a baseline is needed to determine how much of the calculated electric current could be due to numerical effects and/or errors. First, the propagated error for the electric current $\delta I$ is calculated using the corresponding errors $\delta B_{t}$ and $\delta B_{p}$. The numerical effects are represented by the ``current for the potential field'' $I_{pot}$. This is done by calculating the vector of the current-free (potential) magnetic field via extrapolation \citep{alissandrakis81} and applying Ampere's law on the potential magnetic field. The resulting electric current should be equal to zero but this is not the case due to numerical effects. Magnetic partitions are considered non-neutralized only if $I>5 \times I_{pot}$ and $I > 3\times \delta I$. 

For each HARP cutout the method essentially produces a set of electric-current non-neutralized magnetic partitions, along with their corresponding net electric currents $I_{NN}$. From this distribution, the total unsigned non-neutralized electric current $I_{NN,tot}$ = $\Sigma$$|I_{NN}|$ and the maximum unsigned non-neutralized electric current $I_{NN,max} = max(|I_{NN}|)$ of the cut-out can be calculated \citep{kontogiannis17}. For a series of HARP cutouts of a given active region, time series of $I_{NN,tot}$ and $I_{NN,max}$ were constructed, reflecting the evolution of the net electric currents in this region. Additionally, from the maps of the non-neutralized partitions the unsigned magnetic flux pertaining to those non-neutralized partitions $\Phi_{NN,tot}$ can be calculated.


Non-neutralized partitions are found along PILs \citep{geo_titov_mikic12}, but also during flux emergence events. These processes, i.e., emergence and PIL formation will be further discussed in Section~\ref{sec:results}, through a comparative analysis of $I_{NN,tot}$, $I_{NN,max}$, and $\Phi_{NN}$ time series, along with those of the 
total electric current $I_z$ (namely, the surface integral of the electric current density $J_z$)
and the total unsigned magnetic flux, $\Phi$, provided by the Space Weather HARP data \citep[SHARP;][]{bobra14}, for each HARP cut-out.  

We note in passing that this method 
could potentially be used to study individual partitions, too, although we do not attempt this here. Tracking individual partitions should also account for potential merging or splitting effects \citep[see e.g.,][]{2021A&A...647A.146S}. Instead, our focus in this work is to characterize active regions as a whole.

\subsection{Automatic detection of injection events}

An automated detection process was implemented to locate ascending branches of conspicuous peaks in $I_{NN,tot}$, which signify events of net electric current injection in the corresponding time series. To remove potential high frequency noise, the time series were smoothed with a running 6-hour window, corresponding to 31 points (as per the 12 minute cadence of the HMI vector magnetograms). Then the local minima and maxima along with their positions were located. The injection events are segments of notable and persistent increase in $I_{NN,tot}$, and it is assumed that they are demarcated by a local maximum and the local minimum that precedes it. The residual high-frequency small-amplitude variability can cause the detection of several consecutive, short segments along one injection event. A proximity criterion was implemented to merge consecutive segments, \textit{i.e.}, for which the beginning of the following segment is found less than 1 hour after the end of the preceding. To avoid the merging of segments along decreasing parts of the time series, we  demanded that the following segment had a higher average $I_{NN,tot}$ than the preceding. Finally, to avoid variations within the error margin, segments resulting in an increase of less than 10\% were discarded. An example is illustrated in Figure~\ref{fig:method}c, for active region NOAA\,11327. As per the imposed criteria, fairly isolated conspicuous events are always detected but consecutive weak events could be missed because they could be merged leading either to decreasing or marginal, within the error, trend. This could produce a slight underestimation of weak events, which does not, however, alter the conclusions of our study.

\begin{figure*}[htp!]
\centering
\includegraphics[width=18cm]{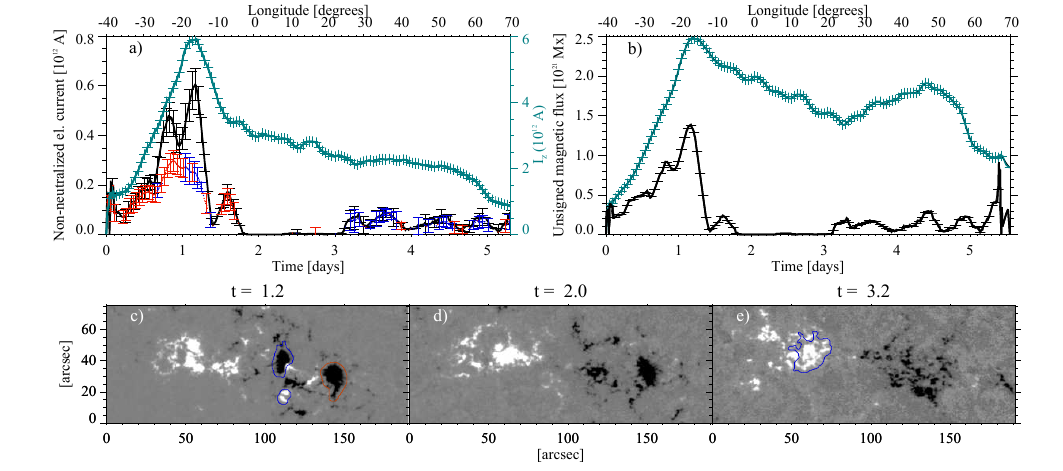}
\caption{a) Time series of $I_{NN,tot}$ (black solid), $I_{NN,max}$ (blue for negative/red for positive) and $I_{z}$ (petrol) for active region NOAA\,11551. b) The total unsigned magnetic flux $\Phi$ (petrol solid) and the total unsigned magnetic flux of the non-neutralized partitions $\Phi_{NN,tot}$ (black). The upper x-axis denotes central meridian distance in both a and b. c-e) Maps of the vertical component $B_{r}$ of the photospheric magnetic field at various instances. Blue and red contours indicate non-neutralized partitions carrying negative and positive electric current, correspondingly. The time is in decimal days since the first magnetogram of the series, taken on 20 August 2012, at 03:46\,UT ($t=0$).}
\label{fig:noaa11551}
\end{figure*}

For the detected time series segments we determined the duration and the average injection rate using a linear fit. For the same segments we calculated the corresponding rates of the total unsigned magnetic flux $\Phi$ and the total unsigned flux of the non-neutralized partitions $\Phi_{NN}$.

\section{Results}\label
{sec:results}
\label{sec:results}
\subsection{Example cases}

\subsubsection{Example of a $\beta$-class active region: NOAA\,11551}\label{noaa11151}

Active region NOAA\,11551 (Figure~\ref{fig:noaa11551}) started emerging on 20 August 2012, at 03:46\,UT and evolved up to a $\beta$-type region. No flares were associated with the region during its passage along the Earth-facing solar hemisphere. Both $I_{z}$ and $\Phi$  (petrol lines in panels (a) and (b), respectively) increased sharply within the first day of emergence and then started to decay. The decrease of $I_{z}$ was sharp up to well into the second day, when the two polarities became fully separated and then it started decreasing slowly until the region decayed and rotated out of view. The magnetic flux exhibited an increase after the third day, during which the decrease of the electric current stagnated. After $t=4.5$\,d both the magnetic flux and the electric current decreased as the region moved past the 60\degr\ heliographic longitude. At such high central meridian distances the measurement of the magnetic field becomes less reliable due to projection and foreshortening effects, as well as the increased noise pattern appearing towards the limb \citep{2014SoPh..289.3483H}. The latter 
are known effects introducing an asymmetric bias in the values of the total unsigned magnetic flux, leading to higher, on average, values towards the western limb \citep[see, e.g., Figure 6 of][]{kontogiannis18}.

\begin{figure*}[htp!]
\centering
\includegraphics[width=18cm]{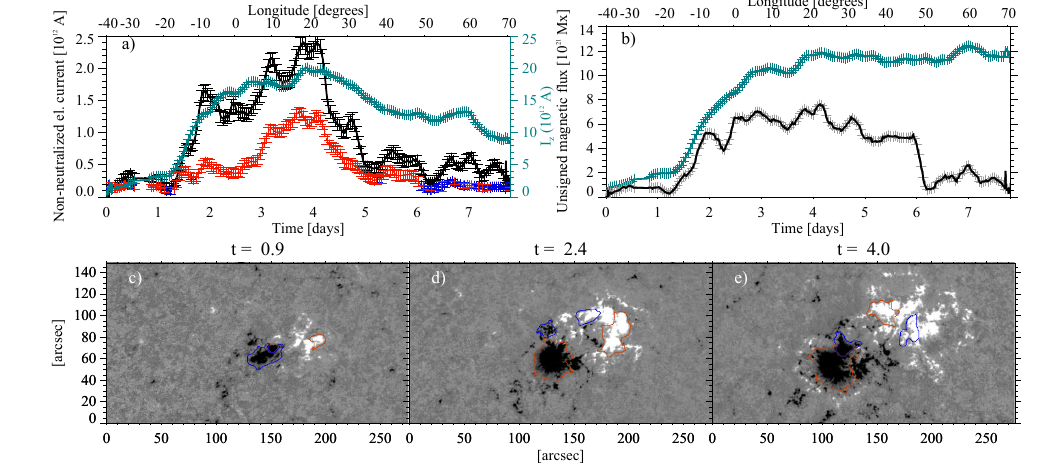}
\caption{Same as Figure~\ref{fig:noaa11551} for active region NOAA\,11327. The first magnetogram of the series was taken on 18 October 2011, at 23:58\,UT ($t=0$).}
\label{fig:noaa11327}
\end{figure*}

The values of $I_{NN,tot}$ were around one order of magnitude lower than $I_{z}$ (note the different y-axes in Figure~\ref{fig:noaa11551}a). Although the overall evolution of $I_{NN,tot}$, $I_{NN,max}$ and $\Phi_{NN}$ followed the general trend of the total unsigned magnetic flux and unsigned total current, there were also notable differences. The total non-neutralized electric current (black line in Figure~\ref{fig:noaa11551}a) also increased during the first day, but exhibited a richer structure. The initial increase was followed by two sharp peaks, reaching up to $5 \times 10^{11}$ and $6 \times 10^{11}$\,A, as well as a lower one seen during the decay. During the most pronounced peaks, $I_{NN,max}$ was always significantly lower than $I_{NN,tot}$ revealing the existence of more than one non-neutralized partitions \citep{kontogiannis17}, which developed during the first day, as the magnetic field emerged and opposite polarities started interacting. The sign of the $I_{NN,max}$ changed several times indicating that during the evolution of the region different non-neutralized partitions, with alternating signs of electric current, were taking the lead. The photospheric magnetogram in Figure~\ref{fig:noaa11551}c shows three non-neutralized partitions of neighboring opposite polarity magnetic flux. During the third day, when $I_{NN,tot}$ dropped to almost zero, the polarities were well separated (Figure~\ref{fig:noaa11551}d). After the end of the third day, $I_{NN,tot}$ and $I_{NN,max}$ increased again, similarly to $I_{z}$, exhibiting a series of peaks which reached up to $10^{11}$\,A, until the region rotated out of view. At all times the $\Phi_{NN}$ followed closely $I_{NN,tot}$; despite being strongly correlated (Pearson correlation at 0.95), the two time series were not identical as seen e.g., by the different relative size of the two peaks around the end of the first day.

\subsubsection{Example of a $\beta\gamma$-class active region: NOAA\,11327}\label{noaa11327}
Active region NOAA\,11327 (Figure~\ref{fig:noaa11327}) started emerging on 18 October 2011, at 23:58\,UT, evolved up to a $\beta\gamma$-type region, but produced no flares during its passage on the solar disk. The magnetic flux and the vertical electric current increased gradually during the first day of emergence and then increased precipitously during the following day reaching a maximum around the end of the fourth day. Then, $I_{z}$ started to decrease slowly for as long as the region was observed, while $\Phi$ remained relatively constant, with a slight tendency to increase.   

The $I_{NN,tot}$ was, overall, one order of magnitude lower than $I_{z}$ and exhibited more structure, with discrete peaks and a more drastic decrease after the maximum at the end of the fourth day. There was a broad, low peak during the first day of emergence, when the two polarities were still close to each other, but $I_{NN,tot}$ decreased briefly as the dipole separated. It then increased sharply during the bulk of the emergence, during which more non-neutralized partitions appeared in close proximity (see also Figure~\ref{fig:noaa11327}d). In this case, however, the $I_{NN,max}$ did not change sign, which means that the corresponding non-neutralized partitions  always carried positive electric current, as a result of one compact negative-flux, positive-current partition having formed (Figure~\ref{fig:noaa11327}d). Sign changes were observed after the end of the sixth day, when the net current in the active region had decreased considerably and the magnetic flux was scattered into several smaller partitions. 

The total unsigned magnetic flux of the non-neutralized partitions, $\Phi_{NN}$, followed a similar trend as the $I_{NN,tot}$, with one notable difference: although $I_{NN,tot}$ exhibited sharp peaks at t = 2, 3 and 4 d, as well as a drastic decrease after the fourth day, the corresponding peaks in magnetic flux were not so pronounced and the overall variation of $\Phi_{NN}$ was  relatively subtle until the end of the sixth day. This clearly implies that interaction between magnetic partitions can change the injected electric current significantly without notable changes in their magnetic flux.

\begin{figure*}[htp!]
\centering
\includegraphics[width=18cm]{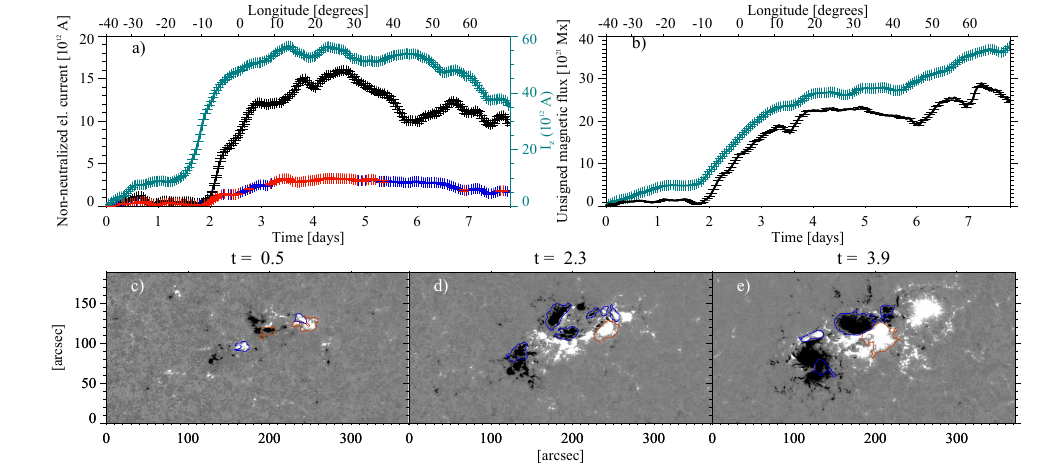}
\caption{Same as Figure~\ref{fig:noaa11551} but for active region NOAA\,11158. The first magnetogram of the series was taken on 10 February 2011, at 21:58\,UT ($t=0$).}
\label{fig:noaa11158}
\end{figure*}

\subsubsection{Example of a $\beta\gamma\delta$-class active region: NOAA\,11158}\label{noaa11158}

Active region NOAA\,11158 (Figure~\ref{fig:noaa11158}) started emerging on 10 February 2011, at 21:58\,UT and developed into a $\beta\gamma\delta$-type region. It produced the first X-class flare of Solar Cycle 24 and, having been closely monitored by the SDO, it was very well-studied \citep[see, e.g.,][and references therein]{tziotziou13,inoue14,2015RAA....15.1547V}.

During the first two days of evolution two bipoles emerged, a smaller one at the 
south-east (lower left)
and a bigger one at the 
north-west (upper right). 
This led to a slow increase of the magnetic flux (Figure~\ref{fig:noaa11158}b), during most of which, there was also an increase in $I_{z}$ and $I_{NN,tot}$ (Figure~\ref{fig:noaa11158}a). Initially, when the two polarities of each dipole started to separate, $I_{z}$ remained almost constant, following closely the evolution of $\Phi$, while $I_{NN,tot}$ decreased until the end of the second day. \citet{2020ApJ...893..123A} report a similar decrease in the degree of current neutralization. Our results suggest that the observed decrease of $I_{NN,tot}$ is a consequence of the separation of the dipoles.

\begin{figure*}[htp!]
\centering
\includegraphics[width=\hsize]{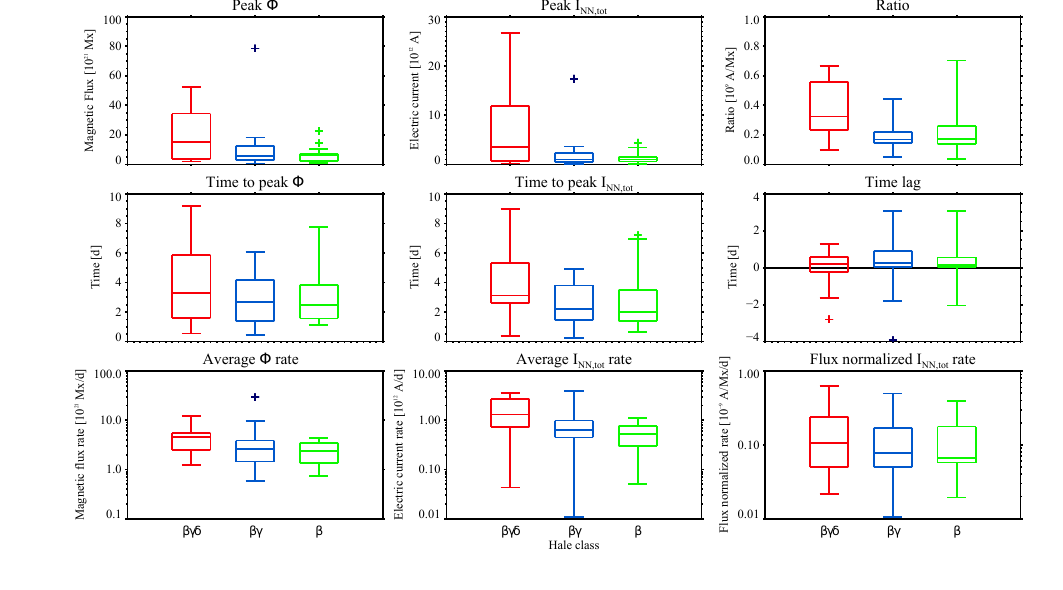}
\caption{Box-and-whiskers plots of time series characteristics for the regions of Table \ref{table:1}. 
The lower and upper boundaries of the box indicate the 25th and 75th percentile, respectively, the horizontal line drawn inside the box indicates the median and the whiskers indicate the minimum and maximum value for each sample. All percentiles are calculated excluding outliers, which are found 3$\sigma$ (the standard deviation of the sample) away from the median and are marked with crosses. The top row contains the peak values of $\Phi$ (left), $I_{NN,tot}$ (middle) and their ratio (right). The middle row contains the time when the corresponding peak values occurred (left and middle) as well as their time lag (right), with positive lag meaning that the peak in $I_{NN,tot}$ occurred first. The bottom row contains the average magnetic flux rate (left), the average $I_{NN,tot}$ rate (middle) and the flux-normalized average $I_{NN,tot}$ rate. Green, blue and red indicate $\beta$-, $\beta\gamma$- and $\beta\gamma\delta$-regions, respectively. All quantities were calculated for central meridian distances within $\pm$45\degr.}
\label{fig:timeseries_stats}
\end{figure*}

At the end of the second day the evolution of NOAA\,11158 became dramatic, with massive emergence of magnetic flux, accompanied by intense flaring activity, comprising several C- and M-class flares, as well as one X-class flare, at t = 4.15\,d. The total unsigned flux increased almost twice as fast in comparison to the first two days, it reached a local maximum around t = 3.4\,d and then increased with a much lower rate until t = 5.3\,d. The flux further increased after t = 6\,d. The increase of the electric current after the end of the second day was sharper, with both $I_{z}$ and $I_{NN,tot}$ reaching local maxima at t = 2.9\,d. Then the two parameters followed a slightly different evolution, $I_{z}$ reaching a maximum at t = 4\,d and $I_{NN,tot}$ reaching a maximum half a day later, at t = 4.5\,d. As the active region evolved further and entered its decay phase, both parameters decreased slowly, with $I_{NN,tot}$ exhibiting a faster decrease. The $I_{NN,max}$ followed the increase of $I_{NN,tot}$, but reached a plateau after the end of the third day, switching sign three times, as indicated by the different colors in Figure~\ref{fig:noaa11158}a. Since the flux was almost balanced during this phase (not shown) and the magnetic flux carried by the non-neutralized partitions, $\Phi_{NN}$, also varied slowly (Figure~\ref{fig:noaa11158}b), this is interpreted as different magnetic partitions acquiring higher non-neutralized electric current as a result of their interaction. In NOAA\,11158, $I_{NN,tot}$ is not an order of magnitude but only about 3-4 times lower than $I_{z}$, indicating that a significantly higher part of the region is carrying non-neutralized electric currents. Additionally, the $I_{NN,tot}$ is around five times higher than $I_{NN,max}$ indicating that many non-neutralized partitions along the PIL contribute to the total net current \citep[see also Figure\,3 in][]{kontogiannis17}. Strongly sheared and complicated PILs are typical characteristics of $\delta$-spot regions, denoting a current-carrying magnetic structure in the regions' core.

\startlongtable
\begin{deluxetable*}{cccccccc}
\tabletypesize{\scriptsize}
\tablecaption{Table 2. Median values of the time series properties presented in Figure~\ref{fig:timeseries_stats}.}
\tablenum{2}
\tablehead{\colhead{Hale Class} & \colhead{Peak $\Phi$} & \colhead{Peak $I_{NN,tot}$} & Peak $I_{NN,tot}$/$\Phi$ & \colhead{Peak time $\Phi$} & \colhead{Peak time $I_{NN,tot}$} & \colhead{$\Phi$ rate} & \colhead{$I_{NN,tot}$ rate} \\ 
\colhead{} & \colhead{$10^{21}\,Mx$} & \colhead{$10^{12}\,A$} & $10^{9}\,AMx^{-1}$ & \colhead{days} & \colhead{days} & \colhead{$10^{21}\,Mxd^{-1}$} & \colhead{$10^{12}\,Ad^{-1}$} } 
\startdata
       $\beta$ &  6.0  &  1.0 & 0.17 &  2.5  &  2.0  &  2.4 & 0.5 \\
       $\beta\gamma$ &  5.9  &  1.0 & 0.17 &  2.7 &  2.2  & 2.6 & 0.6 \\
       $\beta\gamma\delta$ &  15.4  &   3.6 & 0.3 & 3.3  & 3.1  & 4.5 &  1.3 \\
\enddata
\label{table:2}
\end{deluxetable*}

\subsection{Temporal behavior of non-neutralized electric currents}

\subsubsection{The temporal evolution of $I_{NN,tot}$}

To generalize the previous discussion on  the full sample of active regions contained in Table~\ref{table:1}, we calculated some synoptic characteristics of the time series, such as the peak values of $\Phi$ and $I_{NN,tot}$, their occurrence times, as well as the corresponding average time lags between the peaks. These are given in decimal days from the start of the emergence and provide a measure of how fast the regions evolve up to their maximum magnetic flux and electric current. Dividing the peak values by the occurrence times, rudimentary $\Phi$ and $I_{NN,tot}$ variation rates were calculated. Figure~\ref{fig:timeseries_stats} and Table~\ref{table:2} summarize these  results. 

The different magnetic complexity of active region classes $\beta$, $\beta\gamma$ and $\beta\gamma\delta$ is reflected in their total unsigned magnetic flux and electric current. As seen in the upper row of Figure~\ref{fig:timeseries_stats}, $\delta$-regions clearly stand out in terms of maximum $\Phi$ and $I_{NN,tot}$, with median values being more than three times higher than the median $I_{NN,tot}$ values for $\beta$ and $\beta\gamma$-regions. We note in passing that the four, non-outlier $\delta$-spot regions with the lowest magnetic flux and non-neutralized electric current, found in the typical range for $\beta$-regions, produced none (NOAA\,11561, 11431 and 11273) or only two C-class flares (NOAA\,11645). From the $\beta\gamma$-regions (Figure~\ref{fig:timeseries_stats}, upper row, blue boxes), the outlier is NOAA\,12339, which in terms of flare productivity is truly comparable to $\delta$-regions, having produced many C- and M- as well as one X-class flare. Similarly, the two outliers from the $\beta$-regions, namely NOAA\,11130 and 11460, were also the most flare-prolific $\beta$-regions of the sample. These findings agree with the results of \citet{kontogiannis17}, where a correlation was found between the flare index \citep{2005ApJ...629.1141A} and the $I_{NN,tot}$, for a limited sample of regions. In fact the Pearson correlation coefficient between the peak $I_{NN,tot}$, $I_{NN,max}$ and the flare index for our sample is 0.82 and 0.87. A more thorough investigation on the connection with imminent eruptive activity is planned for the future.

The ratio of the peak $I_{NN,tot}$ over the peak $\Phi$ for $\delta$-spot regions is almost twice as high as that of the rest of the sample. This is strong indication that the excess net current in these regions is not due to their larger size but due to their complexity. One may not completely rule out the “big region syndrome” implying large values of non-potentiality metrics in large active regions for extensive properties, albeit net currents are inherently intensive properties, relying on internal structure and complexity. It is also possible that these regions have intrinsically higher electric current upon emergence. 

The middle row of Figure~\ref{fig:timeseries_stats} shows the absolute and relative timing of the $\Phi$ and $I_{NN,tot}$ peak values, that is, the time it takes for active regions to reach their maximum extent and net current content. For $\delta$-spot regions, these times are longer, with NOAA\,11515 and 12673 exhibiting the highest ones. These extreme active regions evolved over many days, during which new flux emerged near already established flux systems \citep{2014A&A...562A.110L,2018A&A...612A.101V}. From the $\beta$-regions sample, NOAA\,11460 is an outlier in terms of peak $I_{NN,tot}$ time. Overall, the peak values in $\Phi$ and $I_{NN,tot}$ are not synchronized. In most cases this difference is in the order few hours and the time lags are predominantly positive, which means that the peak in $I_{NN,tot}$ occurs earlier than the one in $\Phi$. In some cases, magnetic field interactions or flux emergence in later stages of evolution can lead to the peak $I_{NN,tot}$ value to occur later than the $\Phi$ peak. This is also the case for the two outliers, NOAA\,11273 and 11510, a $\beta\gamma\delta$ and a $\beta\gamma$ region, respectively. Therefore, the intrinsic details of flux emergence and subsequent photospheric motions determine the relative timing of magnetic flux and non-neutralized electric current.

Although it takes longer for the $\delta$-spot regions to reach their peak $\Phi$ and $I_{NN,tot}$ values, these are much higher than those measured for the other complexity classes. Hence,  
$\delta$-spot regions exhibit the highest rates (Figure~\ref{fig:timeseries_stats}, bottom row). The median value for the magnetic flux rates (see Table~\ref{table:2}) is within the range of previous results by \citet{2011PASJ...63.1047O}, \citet{2017ApJ...842....3N}, and \citet{2019MNRAS.484.4393K}, if converted to $Mxh^{-1}$. The flare-prolific $\beta\gamma$-region NOAA\,12339 is an outlier also in terms of magnetic flux rate. The median rate of $I_{NN,tot}$ for $\delta$ regions is twice as high as the ones measured for $\beta$ and $\beta\gamma$ ones. The flux-normalized increase rates of $I_{NN,tot}$ are still higher for the most complex regions, which indicates that net current rates are not directly proportional to the magnetic flux (and thus the size) of active regions.

\subsubsection{The temporal evolution of $I_{NN,max}$}

\begin{figure*}[htp!]
\centering
\includegraphics[width=\hsize]{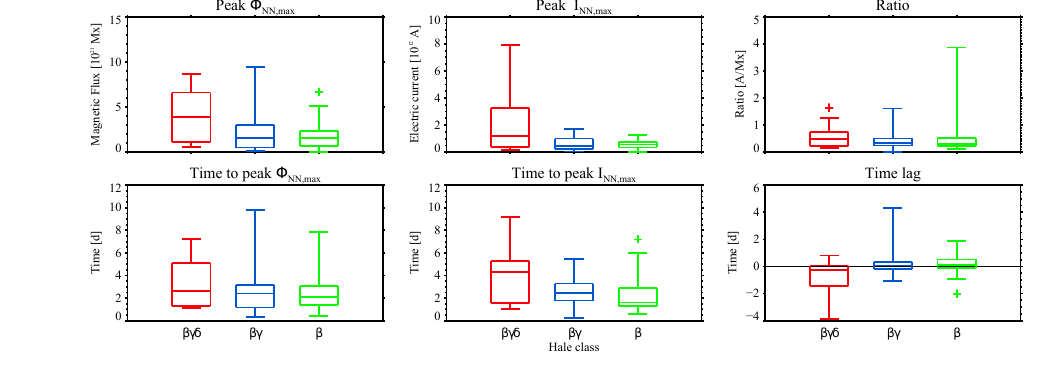}
\caption{Same as Figure~\ref{fig:timeseries_stats}, but for the time series of $I_{NN,max}$ and the associated magnetic flux $\Phi_{NN,max}$. The top row contains the peak values of $\Phi_{NN,max}$ (left), $I_{NN,max}$ (middle) and their ratio (right). The bottom row contains the time when the corresponding peak values occurred (left and middle) as well as their time lag (right), with positive lag meaning that the peak in $I_{NN,max}$ occurred first. All the characteristics were calculated for central meridian distances lower than 45\degr\;EW.}
\label{fig:timeseries_stats_nnmax}
\end{figure*}

While the time series of $I_{NN,tot}$ show how the total amount of non-neutralized electric currents evolves, the corresponding time series of $I_{NN,max}$ show how the electric current carried by the most strongly non-neutralized partition at each instance changes over time. As different parts of the region emerge and evolve, the partition carrying $I_{NN,max}$ can change as well, as does the associated magnetic flux $\Phi_{NN,max}$. This is why, as shown in the descriptions of Figures~\ref{fig:noaa11551},~\ref{fig:noaa11327} and \ref{fig:noaa11158}, the sign of the $I_{NN,max}$ may change during its evolution. In Figure~\ref{fig:timeseries_stats_nnmax}, the corresponding time series of the sample are quantified, similarly to the preceding analysis, in order to find differences between the different complexity classes.

As seen in the panels of the first row of Figure~\ref{fig:timeseries_stats_nnmax}, $\delta$-spot regions are associated with the highest $I_{NN,max}$ and $\Phi_{NN,max}$ values. This means that as these regions evolve, their individual partitions develop the highest electric currents in comparison to the simpler ones. The ratio of the highest individual electric current $I_{NN,max}$ over the corresponding magnetic flux $\Phi_{NN,max}$ is statistically higher as the complexity of the region increases (NOAA\,11682 is an outlier). These panels show that for the most complex regions, the increase in the net current is not proportional to the increase in the corresponding flux carried by the magnetic partitions. 

The panels at the bottom row of Figure~\ref{fig:timeseries_stats_nnmax} show the time required for the peak value of $I_{NN,max}$ and $\Phi_{NN,max}$ to develop in the examined sample of active regions. Similarly to what was demonstrated in Figure~\ref{fig:timeseries_stats}, $\delta$-spot regions take longer to evolve in comparison to $\beta\gamma$ and $\beta$ regions. However, here the distinction between the three classes in terms of the time-lag between the peak values in $I_{NN,max}$ and $\Phi_{NN,max}$ is clearer. Notably, active region NOAA\,11460, a flare-prolific $\beta$-region is again an outlier, demonstrating longer $I_{NN,max}$ peak time and shorter time lag, as well as higher peak $\Phi_{NN,max}$. 
In fact, the majority of $\delta$-spot regions exhibit negative lag, meaning that the peak value of $\Phi_{NN,max}$ occurred earlier. This is another, clear indication that the net electric current continues to increase in non-neutralized partitions after they have fully emerged and reached their peak magnetic flux. 

\begin{figure}[ t!]
\centering
\includegraphics[width=8cm]{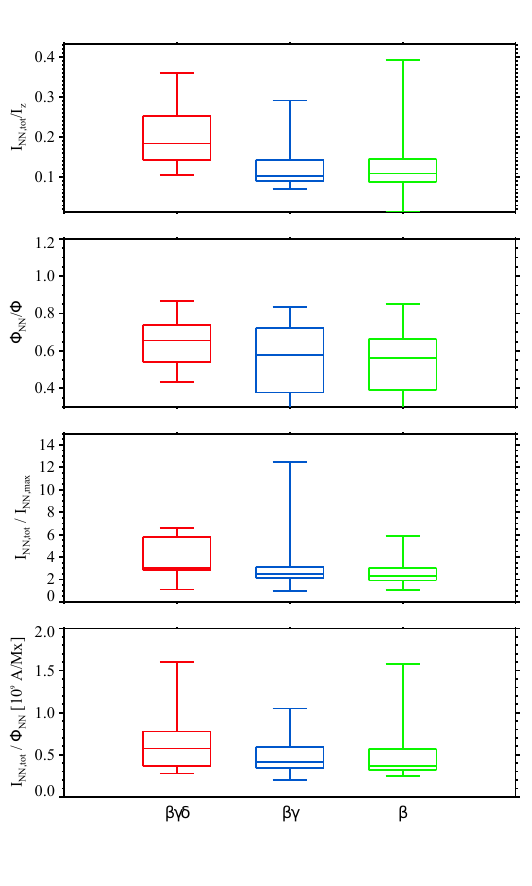}
\caption{From top to bottom: box-and-whiskers plots of the ratios $I_{NN,tot}/I_{z}$, $\Phi_{NN}/\Phi$, $I_{NN,tot}/I_{NN,max}$, and $I_{NN,tot}/\Phi_{NN}$, calculated at the time of peak $I_{NN,tot}$.}
\label{fig:regions_ratios}
\end{figure}

\subsection{The relation between $I_{NN,tot}$, $I_{NN,max}$ and $I_{z}$}

The three active regions presented in Figures~\ref{fig:noaa11551}, \ref{fig:noaa11327} and \ref{fig:noaa11158} differ not only in the amount of non-neutralized electric currents they contained, but also in the ratio of this electric current over $I_{z}$. The same is also true for $\Phi_{NN}$, \textit{i.e.}, the magnetic flux associated with $I_{NN,tot}$. 
Given that these quantities are the corresponding parts of $I_{z}$ and $\Phi$ that pertain to the non-neutralized current part of the regions, it could be surmised that the part of an active region that carries significant net currents increases with magnetic complexity, which is highest for NOAA\,11158 and lowest for NOAA\,11551. For these regions the corresponding ratios were even lower also when the two polarities were well separated and had increased when magnetic flux was emerging or when opposite polarities were in close proximity. 

\begin{figure*}[htp!]
\centering
\includegraphics[width=18cm]{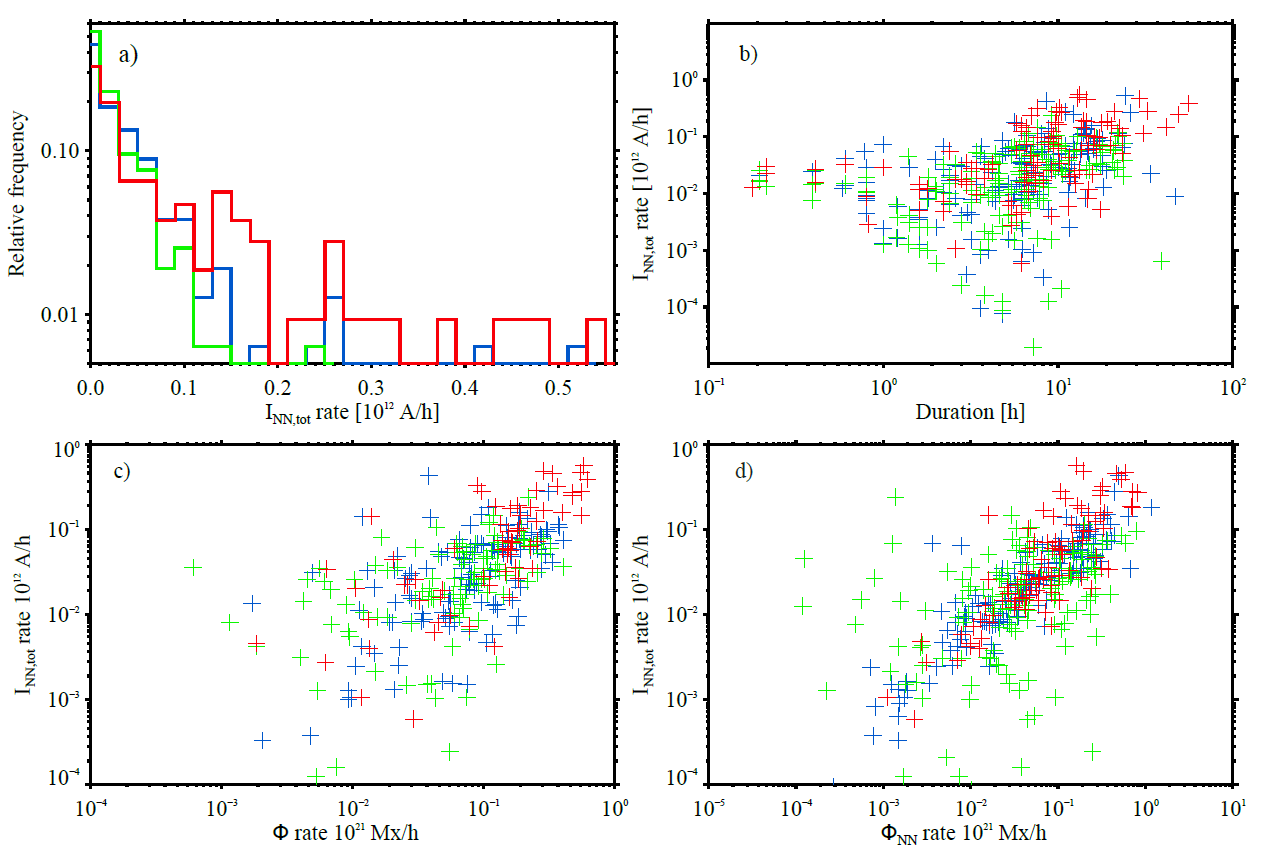}
\caption{Electric-current injection event  statistics for the active regions of Table~\ref{table:1}. a) Histograms of injection rates, b) $I_{NN,tot}$ rate versus injection duration, c) $I_{NN,tot}$ rate vs. magnetic flux rate , and d) $I_{NN,tot}$ rate vs. magnetic flux rate of the non-neutralized partitions. In all panels, green, blue and red colors indicate statistics for $\beta$-, $\beta\gamma$- and $\beta\gamma\delta$-type regions, correspondingly.}
\label{fig:inj_events}
\end{figure*}

Figure~\ref{fig:regions_ratios} extends these findings to the entire sample, showing the distribution of the non-neutralized portion of the regions, in terms of electric current and magnetic flux. These ratios were calculated at the times of maximum $I_{NN,tot}$, for the smoothed time series and for central meridian distances within $\pm$45\degr. As seen in the upper panel, $\delta$-spot regions exhibit twice as high $I_{NN,tot}/I_{z}$ ratios compared  to $\beta$ and $\beta\gamma$ regions, the bulk of which are close to 0.1. The ratio $\Phi_{NN}/\Phi$ exhibits smaller differences per complexity class, with a median around 0.65 for $\delta$-spot regions. The higher ratios and smaller separation between the different classes could be a side effect of the calculation method, because the entire flux contained in a non-neutralized partition  is considered non-neutralized although the net current may be occurring from a fraction of this partition (see, e.g., Fig~\ref{fig:method}c). Thus, magnetic flux associated with the current-carrying magnetic field could be overestimated. Nevertheless, $\delta$-spot regions are located at the upper range of the $\Phi_{NN}/\Phi$ ratio, while $\beta$ and $\beta\gamma$ largely overlap each other. The ratio of non-neutralized electric current is not significantly correlated with the maximum magnetic flux of the region (Pearson correlation coefficient 0.2 for $\delta$-regions). Therefore, the observed differences between different complexity classes do not reflect their total magnetic flux content. 

Another interesting finding pertains to the ratio $I_{NN,tot}/I_{NN,max}$. For the bulk of $\beta$ and $\beta\gamma$ regions it is less than 3, meaning that the $I_{NN,tot}$ is barely twice as high as $I_{NN,max}$, \textit{i.e.}, the electric current carried by the most non-neutralized partition. Conversely, for $\delta$ regions the corresponding ratio acquires higher values, indicating the existence of many non-neutralized partitions in close proximity, forming complicated PILs \citep{kontogiannis17,kontogiannis19}.

The ratio between the non-neutralized electric current over the current-carrying magnetic flux is examined at the bottom panel of Figure~\ref{fig:regions_ratios}. No notable differences are seen for $\beta$- and $\beta\gamma$-regions, but the current-carrying flux in $\delta$-regions carries more electric current per unit non-neutralized flux in comparison to regions belonging to the other two classes. This result is in line with the observed highly-twisted structures (flux ropes and/or sheared magnetic arcades) these regions often host \citep[see \textit{e.g.},][and references therein]{2020SSRv..216..131P} and points to a systematic characteristic exhibited by these active regions. 

\subsection{Net electric-current injection events}
\label{sec:inj_ev}

As seen in Figures~\ref{fig:noaa11551}, \ref{fig:noaa11327} and \ref{fig:noaa11158}, peaks in $I_{NN,tot}$ are found in sequence or superposition and indicate injection events and overall build-up of electric currents as a result of both emergence and photospheric interactions. 
To study these events in more detail, the detection algorithm was applied to the  $I_{NN,tot}$ time series of the sample. The statistics for the detected events are summarized in Figure~\ref{fig:inj_events}.

The relative frequency of the injection rates drops as the injection rate increases (Figure~\ref{fig:inj_events}a), but the highest injection rates are found in $\delta$-spot regions, followed by $\beta\gamma$ ones. The former show increased relative frequency of events with rates between 0.1 and $0.5\cdot10^{12}\,Ah^{-1}$, whereas for $\beta$- regions they reach up to $0.25\cdot10^{12}\,Ah^{-1}$.  Overall the injection rate is moderately correlated with the event duration (0.45 Pearson correlation coefficient, Figure~\ref{fig:inj_events}b). This correlation is slightly stronger for $\delta$-regions (0.5) in comparison to the rest (0.38) and the duration of the events is longer for the most complex regions.
   
At the bottom row of Figure~\ref{fig:inj_events} the $I_{NN,tot}$ versus $\Phi$ (panel c) and $\Phi_{NN}$ (panel d) rates of the events are plotted. There is an overall positive correlation between these quantities (Pearson correlation coefficient equal to 0.54 and 0.52, respectively), reflecting the fact that in many cases the increase of the electric current is associated or happens during the emergence of magnetic flux. This is the case seen, e.g., in NOAA\,11158 (Figure~\ref{fig:noaa11158}), during the first day and, mainly, during the third day, where the bulk of magnetic flux emerged and was accompanied by strong interaction between magnetic footpoints. However, the $I_{NN,tot}$ increase is not always associated with increasing magnetic flux since, as indicated by the two scatter plots, there is a considerable number of events associated with very small and even negative magnetic flux rates, indicating flux decay. The correlation coefficients between $I_{NN,tot}$ and $\Phi$ rates for $\delta$- $\beta\gamma$- and $\beta$-regions are 0.66, 0.30 and 0.47. The corresponding correlation coefficient between the $I_{NN,tot}$ and $\Phi_{NN}$ rates (Figure~\ref{fig:inj_events}d) are lower, that is, 0.52, 0.51 and 0.28. For the injection events with rates higher than $0.12\cdot10^{12}\,Ah^{-1}$ (which are mostly events detected in $\delta$-regions, see also Figure~\ref{fig:inj_events}a) the correlation for $\delta$-regions, which exhibit most of these rates, drops to 0.47 and 0.20 for $\Phi$ and $\Phi_{NN}$, respectively. This correlation analysis for the injection events corroborates the conclusions drawn from the preceding analysis of time series properties. The current-carrying magnetic flux $\Phi_{NN}$ peaks earlier than the $I_{NN,tot}$, which keeps increasing (see Figure~\ref{fig:timeseries_stats_nnmax}) and exhibiting injection events, even after $\Phi_{NN}$ is more or less constant (see \textit{e.g.}, Figure~\ref{fig:noaa11158}). 

\section{Discussion and conclusions}\label{sec:conclusions}

This study presents, to the best of our  knowledge, the first statistical study of the evolution of non-neutralized electric currents in a statistically significant sample of solar active regions. For the derivation of non-neutralized electric currents the method of \citet{geo_titov_mikic12} was used, that relies on the calculation of the vertical electric current density and uses criteria on error propagation and numerical errors. The resulting time series of $I_{NN,tot}$ are relevant to the non-neutralized current-carrying part of an active region and exhibit a rich structure, not seen in the time series of the surface integrated total unsigned current, $I_{z}$. Through a comparative study of the $I_{NN,tot}$, $I_{NN,max}$, $\Phi_{NN}$ time series, along with those of $I_{z}$ and $\Phi$, we reach a number of conclusions regarding the evolution of net electric currents in active regions and the differences between regions of different magnetic complexity.

In all types of regions, the unsigned magnetic flux and the electric current exhibit similar (but far from identical) evolutionary trends, in terms of overall increase and decay. However, the peak values of $\Phi$ and $I_{NN,tot}$, representing their maximum size and complexity respectively, are in principle not synchronized, instead, the relative timing of these peaks depends not only on the emergence of magnetic flux but also on the subsequent interactions between magnetic partitions. Thus the build-up of net electric currents is a process driven by magnetic flux emergence but significantly shaped by interactions between flux systems. The latter factor is more pronounced in $\delta$-spot regions, where $I_{NN,tot}$ continues to increase after the magnetic flux $\Phi_{NN}$, associated with the net currents has peaked for most regions. The importance of polarity collision has been highlighted for some cases of $\delta$ regions that were extremely flare-prolific and exhibited various morphological configurations \citep{2018A&A...612A.101V,toriumi17,2019ApJ...871...67C}. The results presented here generalize these case studies, from the point of view of net electric current injection. 

During the flux decay phase, non-neutralized electric currents decrease faster than the magnetic flux, 
implying the vital relation between net electric currents and flux emergence along PILs, which leads to these currents' fast decrease during active region decay and their settling into more relaxed configurations with scattered, well-separated polarities. This decrease is more pronounced in $I_{NN,tot}$ than in $I_{z}$, corroborating the higher relevance of  $I_{NN,tot}$ to magnetic complexity. Even during decay, though, incidental complexity due to interaction can still lead to injection of net electric currents into the corona, as indicated by significant peaks often seen in late stages of active region evolution. 

To further study the injection of electric current in active regions, an automated detection method was implemented and a statistical study was carried out. The more complex regions, i.e., $\beta\gamma$ and $\beta\gamma\delta$ were characterized by injection events with correspondingly higher rate and duration. The calculated correlations between the $I_{NN,tot}$ injection rates with those of $\Phi$ and $\Phi_{NN}$ reflect again the combined role of magnetic flux emergence and subsequent interactions in the formation of the three complexity classes. A future step would involve studying the 
characteristics of injection events on a larger sample of active regions, with no limitation on evolutionary stage, and explore the association with imminent flaring and eruptive activity. 
This might contribute clues toward prediction of flares and eruptions (i.e., CMEs), which is a key challenge in solar weather prediction \citep{2023AdSpR..71.2017K}.

In this work, we extended the findings of \citet{kontogiannis17} regarding the relevance of non-neutralized electric currents to imminent flare productivity using a statistically significant sample of active regions. In several cases, $I_{NN,tot}$ evolution was unrelated to the size of active regions, in terms of their magnetic flux. Such “intensive” behavior is considered highly relevant to CME prediction \citep{2016ApJ...821..127B}. Our results indicate that the relevance of the examined parameters to flare and CME initiation is worth exploring in the future. 

Regarding the relation between non-neutralized electric currents and complexity, it was shown that active regions with $\delta$ spots differ both in terms of peak values and evolution. They contain not only the highest magnetic flux but also the highest net electric currents, even when the latter are normalized to their peak magnetic flux content. The same is true for the increase rate of $I_{NN,tot}$, as well as the ratio $I_{NN,tot}/I_{z}$, \textit{i.e.}, the part of the region that is current-carrying. For lower complexity cases, $\beta$-regions are practically indistinguishable from $\beta\gamma$-regions in terms of those properties. Since the non-neutralized electric currents are directly associated with the presence of sheared PILs and, in our formulation, the ratio of $I_{NN,tot}$ over $\Phi$ or $\Phi_{NN}$ is a proxy of twist, the results presented here signify an inherently higher twist as well as different formation processes for the regions hosting $\delta$-spots.

The study of Hale $\delta$-class active regions in terms of net currents is extremely important both due to their high eruptive potential \citep{2000ApJ...540..583S} and because of our incomplete knowledge on which mechanism drives their formation \citep[see \textit{e.g.},][and references therein]{2003A&A...397..305L,kniznik18}. The approach used in the present study can be used to distinguish and probe the net current-carrying part of an active region. Determining its properties can have implications on the formation mechanism, even more so since typical photospheric proxies that characterize active regions as a whole are not very efficient on distinguishing emerging flux tubes of different sub-photospheric properties \citep{kniznik18}. In this context, \citet{norton22} found that the emergence rate of these regions is almost twice as high (1.9 times) as those of $\beta$, which is also corroborated by the present study (see Table~\ref{table:2}). Here it was additionally shown that the non-neutralized electric currents of $\delta$-spot regions increase 2.6 times faster than in other active region types, with individual injection events exhibiting even higher relative rates (Figure~\ref{fig:inj_events}). The median of the maximum ratio $\Phi_{NN}$/$\Phi$ was 0.65 (Figure~\ref{fig:regions_ratios}), which is directly comparable with the so called ``degree of $\delta$'', which \citet{norton22} defined as the magnetic flux fraction that participates in the $\delta$ configuration and amounts to 0.72 in their study. Recently, \citet{levens23} examined a limited sample of regions and showed that the current helicity, electric current and twist in magnetic knots, that is, the constituents of $\delta$-spots, are all compatible with the kink instability. Therefore, we conclude that the approach used in the present study could be used in the future to contribute to the determination of $\delta$-spots properties, providing more clues on the sub-surface condition in these regions. 

An additional benefit of our methodology is that the derivation of $I_{NN,tot}$ is not based on assumptions concerning the chirality of the active region \citep[as \textit{e.g.}, in][]{2020ApJ...893..123A} and/or specific region selection \citep{2017ApJ...846L...6L,2024ApJ...961..148L}, which are required to determine the so-called degree of current neutralization \citep{torok14}. The presented results are in line with those derived from the study of the degree of current neutralization \citep{2017ApJ...846L...6L,2019MNRAS.486.4936V,2020ApJ...893..123A,2024ApJ...961..148L,2020ApJ...895...18B,2023ApJ...943...80W,2020ApJ...900...38H}. Another future step would be a comparative analysis of the evolution of active regions using both approaches, namely the degree of current neutralization along with the ratio $I_{NN,tot}/I_{z}$ (or its complement $1 - I_{NN,tot}/I_{z}$) and the neutrality factor proposed by \citet{geo_titov_mikic12}. 

\begin{acknowledgements}
We would like to thank the anonymous referee for providing useful comments, which improved the content and presentation of this study. This work was supported by grant KO~6283/2-1 of the Deutsche Forschungsgemeinschaft (DFG). For this paper, data from SDO/AIA and SDO/HMI were used. These are courtesy of NASA/SDO and the AIA, EVE, and HMI science teams and are publicly available through the Joint Science Operations Center at the \url{jsoc.stanford.edu}. Some of the calculations were performed using source codes made publicly available by the H2020 project FLARECAST (\url{https://dev.flarecast.eu/stash/projects/}).
\end{acknowledgements}
\
\
\ 
\ 

\facilities{SDO}


\software{SolarSoft IDL \citep{bentley98,Freeland98}, FLARECAST \citep{2021JSWSC..11...39G}  }

\bibliography{references}{}

\begin{thebibliography}{}
\expandafter\ifx\csname natexlab\endcsname\relax\def\natexlab#1{#1}\fi
\providecommand{\url}[1]{\href{#1}{#1}}
\providecommand{\dodoi}[1]{doi:~\href{http://doi.org/#1}{\nolinkurl{#1}}}
\providecommand{\doeprint}[1]{\href{http://ascl.net/#1}{\nolinkurl{http://ascl.net/#1}}}
\providecommand{\doarXiv}[1]{\href{https://arxiv.org/abs/#1}{\nolinkurl{https://arxiv.org/abs/#1}}}

\bibitem[{{Abramenko}(2005)}]{2005ApJ...629.1141A}
{Abramenko}, V.~I. 2005, \apj, 629, 1141, \dodoi{10.1086/431732}

\bibitem[{{Alfv{\'e}n} \& {Carlqvist}(1967)}]{1967SoPh....1..220A}
{Alfv{\'e}n}, H., \& {Carlqvist}, P. 1967, \solphys, 1, 220,
  \dodoi{10.1007/BF00150857}

\bibitem[{{Alissandrakis}(1981)}]{alissandrakis81}
{Alissandrakis}, C.~E. 1981, \aap, 100, 197

\bibitem[{{Angryk} {et~al.}(2020){Angryk}, {Martens}, {Aydin}, {Kempton},
  {Mahajan}, {Basodi}, {Ahmadzadeh}, {Cai}, {Filali Boubrahimi}, {Hamdi},
  {Schuh}, \& {Georgoulis}}]{2020NatSD...7..227A}
{Angryk}, R.~A., {Martens}, P.~C., {Aydin}, B., {et~al.} 2020, Scientific Data,
  7, 227, \dodoi{10.1038/s41597-020-0548-x}

\bibitem[{{Aulanier} {et~al.}(2013){Aulanier}, {D{\'e}moulin}, {Schrijver},
  {Janvier}, {Pariat}, \& {Schmieder}}]{2013A&A...549A..66A}
{Aulanier}, G., {D{\'e}moulin}, P., {Schrijver}, C.~J., {et~al.} 2013, \aap,
  549, A66, \dodoi{10.1051/0004-6361/201220406}

\bibitem[{{Aulanier} {et~al.}(2012){Aulanier}, {Janvier}, \&
  {Schmieder}}]{2012A&A...543A.110A}
{Aulanier}, G., {Janvier}, M., \& {Schmieder}, B. 2012, \aap, 543, A110,
  \dodoi{10.1051/0004-6361/201219311}

\bibitem[{{Avallone} \& {Sun}(2020)}]{2020ApJ...893..123A}
{Avallone}, E.~A., \& {Sun}, X. 2020, \apj, 893, 123,
  \dodoi{10.3847/1538-4357/ab7afa}

\bibitem[{{Barczynski} {et~al.}(2020){Barczynski}, {Aulanier}, {Janvier},
  {Schmieder}, \& {Masson}}]{2020ApJ...895...18B}
{Barczynski}, K., {Aulanier}, G., {Janvier}, M., {Schmieder}, B., \& {Masson},
  S. 2020, \apj, 895, 18, \dodoi{10.3847/1538-4357/ab893d}

\bibitem[{{Barnes} {et~al.}(2005){Barnes}, {Longcope}, \& {Leka}}]{barnes05}
{Barnes}, G., {Longcope}, D.~W., \& {Leka}, K.~D. 2005, \apj, 629, 561,
  \dodoi{10.1086/431175}

\bibitem[{{Bentley} \& {Freeland}(1998)}]{bentley98}
{Bentley}, R.~D., \& {Freeland}, S.~L. 1998, in ESA Special Publication, Vol.
  417, Crossroads for European Solar and Heliospheric Physics. Recent
  Achievements and Future Mission Possibilities, 225--228

\bibitem[{{Bobra} \& {Ilonidis}(2016)}]{2016ApJ...821..127B}
{Bobra}, M.~G., \& {Ilonidis}, S. 2016, \apj, 821, 127,
  \dodoi{10.3847/0004-637X/821/2/127}

\bibitem[{{Bobra} {et~al.}(2014){Bobra}, {Sun}, {Hoeksema}, {Turmon}, {Liu},
  {Hayashi}, {Barnes}, \& {Leka}}]{bobra14}
{Bobra}, M.~G., {Sun}, X., {Hoeksema}, J.~T., {et~al.} 2014, \solphys, 289,
  3549, \dodoi{10.1007/s11207-014-0529-3}

\bibitem[{{Chintzoglou} {et~al.}(2019){Chintzoglou}, {Zhang}, {Cheung}, \&
  {Kazachenko}}]{2019ApJ...871...67C}
{Chintzoglou}, G., {Zhang}, J., {Cheung}, M. C.~M., \& {Kazachenko}, M. 2019,
  \apj, 871, 67, \dodoi{10.3847/1538-4357/aaef30}

\bibitem[{{Dalmasse} {et~al.}(2015){Dalmasse}, {Aulanier}, {D{\'e}moulin},
  {Kliem}, {T{\"o}r{\"o}k}, \& {Pariat}}]{dalmasse15}
{Dalmasse}, K., {Aulanier}, G., {D{\'e}moulin}, P., {et~al.} 2015, \apj, 810,
  17, \dodoi{10.1088/0004-637X/810/1/17}

\bibitem[{{Freeland} \& {Handy}(1998)}]{Freeland98}
{Freeland}, S.~L., \& {Handy}, B.~N. 1998, \solphys, 182, 497,
  \dodoi{10.1023/A:1005038224881}

\bibitem[{{Georgoulis}(2018)}]{2018GMS...235..371G}
{Georgoulis}, M.~K. 2018, in Electric Currents in Geospace and Beyond, ed.
  A.~{Keiling}, O.~{Marghitu}, \& M.~{Wheatland}, Vol. 235, 371--390,
  \dodoi{10.1002/9781119324522.ch22}

\bibitem[{{Georgoulis} {et~al.}(2019){Georgoulis}, {Nindos}, \&
  {Zhang}}]{2019RSPTA.37780094G}
{Georgoulis}, M.~K., {Nindos}, A., \& {Zhang}, H. 2019, Philosophical
  Transactions of the Royal Society of London Series A, 377, 20180094,
  \dodoi{10.1098/rsta.2018.0094}

\bibitem[{{Georgoulis} {et~al.}(2012){Georgoulis}, {Titov}, \&
  {Miki{\'c}}}]{geo_titov_mikic12}
{Georgoulis}, M.~K., {Titov}, V.~S., \& {Miki{\'c}}, Z. 2012, \apj, 761, 61,
  \dodoi{DOI: 10.1088/0004-637X/761/1/61}

\bibitem[{{Georgoulis} {et~al.}(2021){Georgoulis}, {Bloomfield}, {Piana},
  {Massone}, {Soldati}, {Gallagher}, {Pariat}, {Vilmer}, {Buchlin}, {Baudin},
  {Csillaghy}, {Sathiapal}, {Jackson}, {Alingery}, {Benvenuto}, {Campi},
  {Florios}, {Gontikakis}, {Guennou}, {Guerra}, {Kontogiannis}, {Latorre},
  {Murray}, {Park}, {von Stachelski}, {Torbica}, {Vischi}, \&
  {Worsfold}}]{2021JSWSC..11...39G}
{Georgoulis}, M.~K., {Bloomfield}, D.~S., {Piana}, M., {et~al.} 2021, Journal
  of Space Weather and Space Climate, 11, 39, \dodoi{10.1051/swsc/2021023}

\bibitem[{{Hale} {et~al.}(1919){Hale}, {Ellerman}, {Nicholson}, \&
  {Joy}}]{1919ApJ....49..153H}
{Hale}, G.~E., {Ellerman}, F., {Nicholson}, S.~B., \& {Joy}, A.~H. 1919, \apj,
  49, 153, \dodoi{10.1086/142452}

\bibitem[{{He} {et~al.}(2020){He}, {Liu}, {Liu}, {Chen}, {Wang}, \&
  {Wang}}]{2020ApJ...900...38H}
{He}, Y., {Liu}, R., {Liu}, L., {et~al.} 2020, \apj, 900, 38,
  \dodoi{10.3847/1538-4357/aba52a}

\bibitem[{{Hoeksema} {et~al.}(2014){Hoeksema}, {Liu}, {Hayashi}, {Sun},
  {Schou}, {Couvidat}, {Norton}, {Bobra}, {Centeno}, {Leka}, {Barnes}, \&
  {Turmon}}]{2014SoPh..289.3483H}
{Hoeksema}, J.~T., {Liu}, Y., {Hayashi}, K., {et~al.} 2014, \solphys, 289,
  3483, \dodoi{10.1007/s11207-014-0516-8}

\bibitem[{{Inoue} {et~al.}(2014){Inoue}, {Hayashi}, {Magara}, {Choe}, \&
  {Park}}]{inoue14}
{Inoue}, S., {Hayashi}, K., {Magara}, T., {Choe}, G.~S., \& {Park}, Y.~D. 2014,
  \apj, 788, 182, \dodoi{10.1088/0004-637X/788/2/182}

\bibitem[{{Jaeggli} \& {Norton}(2016)}]{2016ApJ...820L..11J}
{Jaeggli}, S.~A., \& {Norton}, A.~A. 2016, \apjl, 820, L11,
  \dodoi{10.3847/2041-8205/820/1/L11}

\bibitem[{{Janvier} {et~al.}(2014){Janvier}, {Aulanier}, {Bommier},
  {Schmieder}, {D{\'e}moulin}, \& {Pariat}}]{janvier14}
{Janvier}, M., {Aulanier}, G., {Bommier}, V., {et~al.} 2014, \apj, 788, 60,
  \dodoi{10.1088/0004-637X/788/1/60}

\bibitem[{{Janvier} {et~al.}(2013){Janvier}, {Aulanier}, {Pariat}, \&
  {D{\'e}moulin}}]{2013A&A...555A..77J}
{Janvier}, M., {Aulanier}, G., {Pariat}, E., \& {D{\'e}moulin}, P. 2013, \aap,
  555, A77, \dodoi{10.1051/0004-6361/201321164}

\bibitem[{{Kazachenko} {et~al.}(2022){Kazachenko}, {Lynch}, {Savcheva}, {Sun},
  \& {Welsch}}]{2022ApJ...926...56K}
{Kazachenko}, M.~D., {Lynch}, B.~J., {Savcheva}, A., {Sun}, X., \& {Welsch},
  B.~T. 2022, \apj, 926, 56, \dodoi{10.3847/1538-4357/ac3af3}

\bibitem[{{Knizhnik} {et~al.}(2018){Knizhnik}, {Linton}, \&
  {DeVore}}]{kniznik18}
{Knizhnik}, K.~J., {Linton}, M.~G., \& {DeVore}, C.~R. 2018, \apj, 864, 89,
  \dodoi{10.3847/1538-4357/aad68c}

\bibitem[{{Kontogiannis}(2023)}]{2023AdSpR..71.2017K}
{Kontogiannis}, I. 2023, Advances in Space Research, 71, 2017,
  \dodoi{10.1016/j.asr.2022.10.008}

\bibitem[{{Kontogiannis} {et~al.}(2019){Kontogiannis}, {Georgoulis}, {Guerra},
  {Park}, \& {Bloomfield}}]{kontogiannis19}
{Kontogiannis}, I., {Georgoulis}, M.~K., {Guerra}, J.~A., {Park}, S.-H., \&
  {Bloomfield}, D.~S. 2019, \solphys, 294, 130, \dodoi{DOI:
  10.1007/s11207-019-1523-6}

\bibitem[{{Kontogiannis} {et~al.}(2017){Kontogiannis}, {Georgoulis}, {Park}, \&
  {Guerra}}]{kontogiannis17}
{Kontogiannis}, I., {Georgoulis}, M.~K., {Park}, S.-H., \& {Guerra}, J.~A.
  2017, \solphys, 292, 159, \dodoi{DOI: 10.1007/s11207-017-1185-1}

\bibitem[{{Kontogiannis} {et~al.}(2018){Kontogiannis}, {Georgoulis}, {Park}, \&
  {Guerra}}]{kontogiannis18}
---. 2018, \solphys, 293, 96, \dodoi{DOI: 10.1007/s11207-018-1317-2}

\bibitem[{{K{\"u}nzel}(1965)}]{1965AN....288..177K}
{K{\"u}nzel}, H. 1965, Astronomische Nachrichten, 288, 177

\bibitem[{{Kutsenko} {et~al.}(2019){Kutsenko}, {Abramenko}, \&
  {Pevtsov}}]{2019MNRAS.484.4393K}
{Kutsenko}, A.~S., {Abramenko}, V.~I., \& {Pevtsov}, A.~A. 2019, \mnras, 484,
  4393, \dodoi{10.1093/mnras/stz308}

\bibitem[{{Leka} {et~al.}(1996){Leka}, {Canfield}, {McClymont}, \& {van
  Driel-Gesztelyi}}]{Leka96}
{Leka}, K.~D., {Canfield}, R.~C., {McClymont}, A.~N., \& {van Driel-Gesztelyi},
  L. 1996, \apj, 462, 547, \dodoi{10.1086/177171}

\bibitem[{{Levens} {et~al.}(2023){Levens}, {Norton}, {Linton}, {Knizhnik}, \&
  {Liu}}]{levens23}
{Levens}, P.~J., {Norton}, A.~A., {Linton}, M.~G., {Knizhnik}, K.~J., \& {Liu},
  Y. 2023, \apjl, 954, L20, \dodoi{10.3847/2041-8213/acf0c6}

\bibitem[{{Liu} {et~al.}(2017){Liu}, {Sun}, {T{\"o}r{\"o}k}, {Titov}, \&
  {Leake}}]{2017ApJ...846L...6L}
{Liu}, Y., {Sun}, X., {T{\"o}r{\"o}k}, T., {Titov}, V.~S., \& {Leake}, J.~E.
  2017, \apjl, 846, L6, \dodoi{10.3847/2041-8213/aa861e}

\bibitem[{{Liu} {et~al.}(2024){Liu}, {T{\"o}r{\"o}k}, {Titov}, {Leake}, {Sun},
  \& {Jin}}]{2024ApJ...961..148L}
{Liu}, Y., {T{\"o}r{\"o}k}, T., {Titov}, V.~S., {et~al.} 2024, \apj, 961, 148,
  \dodoi{10.3847/1538-4357/ad11da}

\bibitem[{{Longcope} \& {Welsch}(2000)}]{longcope_welsch00}
{Longcope}, D.~W., \& {Welsch}, B.~T. 2000, \apj, 545, 1089,
  \dodoi{10.1086/317846}

\bibitem[{{L{\'o}pez Fuentes} {et~al.}(2003){L{\'o}pez Fuentes},
  {D{\'e}moulin}, {Mandrini}, {Pevtsov}, \& {van
  Driel-Gesztelyi}}]{2003A&A...397..305L}
{L{\'o}pez Fuentes}, M.~C., {D{\'e}moulin}, P., {Mandrini}, C.~H., {Pevtsov},
  A.~A., \& {van Driel-Gesztelyi}, L. 2003, \aap, 397, 305,
  \dodoi{10.1051/0004-6361:20021487}

\bibitem[{{Louis} {et~al.}(2014){Louis}, {Puschmann}, {Kliem}, {Balthasar}, \&
  {Denker}}]{2014A&A...562A.110L}
{Louis}, R.~E., {Puschmann}, K.~G., {Kliem}, B., {Balthasar}, H., \& {Denker},
  C. 2014, \aap, 562, A110, \dodoi{10.1051/0004-6361/201321106}

\bibitem[{{McIntosh}(1990)}]{1990SoPh..125..251M}
{McIntosh}, P.~S. 1990, \solphys, 125, 251, \dodoi{10.1007/BF00158405}

\bibitem[{{Mitra} {et~al.}(2020){Mitra}, {Joshi}, \&
  {Prasad}}]{2020SoPh..295...29M}
{Mitra}, P.~K., {Joshi}, B., \& {Prasad}, A. 2020, \solphys, 295, 29,
  \dodoi{10.1007/s11207-020-1596-2}

\bibitem[{{Norton} {et~al.}(2017){Norton}, {Jones}, {Linton}, \&
  {Leake}}]{2017ApJ...842....3N}
{Norton}, A.~A., {Jones}, E.~H., {Linton}, M.~G., \& {Leake}, J.~E. 2017, \apj,
  842, 3, \dodoi{10.3847/1538-4357/aa7052}

\bibitem[{{Norton} {et~al.}(2022){Norton}, {Levens}, {Knizhnik}, {Linton}, \&
  {Liu}}]{norton22}
{Norton}, A.~A., {Levens}, P.~J., {Knizhnik}, K.~J., {Linton}, M.~G., \& {Liu},
  Y. 2022, \apj, 938, 117, \dodoi{10.3847/1538-4357/ac8eb2}

\bibitem[{{Otsuji} {et~al.}(2011){Otsuji}, {Kitai}, {Ichimoto}, \&
  {Shibata}}]{2011PASJ...63.1047O}
{Otsuji}, K., {Kitai}, R., {Ichimoto}, K., \& {Shibata}, K. 2011, \pasj, 63,
  1047, \dodoi{10.1093/pasj/63.5.1047}

\bibitem[{{Parker}(1979)}]{parker79}
{Parker}, E.~N. 1979, {Cosmical magnetic fields: Their origin and their
  activity}

\bibitem[{{Parker}(1996)}]{parker96}
---. 1996, \apj, 471, 485, \dodoi{10.1086/177983}

\bibitem[{{Patsourakos} {et~al.}(2020){Patsourakos}, {Vourlidas},
  {T{\"o}r{\"o}k}, {Kliem}, {Antiochos}, {Archontis}, {Aulanier}, {Cheng},
  {Chintzoglou}, {Georgoulis}, {Green}, {Leake}, {Moore}, {Nindos}, {Syntelis},
  {Yardley}, {Yurchyshyn}, \& {Zhang}}]{2020SSRv..216..131P}
{Patsourakos}, S., {Vourlidas}, A., {T{\"o}r{\"o}k}, T., {et~al.} 2020, \ssr,
  216, 131, \dodoi{10.1007/s11214-020-00757-9}

\bibitem[{{Pesnell} {et~al.}(2012){Pesnell}, {Thompson}, \& {Chamberlin}}]{sdo}
{Pesnell}, W.~D., {Thompson}, B.~J., \& {Chamberlin}, P.~C. 2012, \solphys,
  275, 3, \dodoi{10.1007/s11207-011-9841-3}

\bibitem[{{Sahu} {et~al.}(2023){Sahu}, {Joshi}, {Prasad}, \& {Cho}}]{suraj2023}
{Sahu}, S., {Joshi}, B., {Prasad}, A., \& {Cho}, K.-S. 2023, \apj, 943, 70,
  \dodoi{10.3847/1538-4357/acac2d}

\bibitem[{{Sammis} {et~al.}(2000){Sammis}, {Tang}, \&
  {Zirin}}]{2000ApJ...540..583S}
{Sammis}, I., {Tang}, F., \& {Zirin}, H. 2000, \apj, 540, 583,
  \dodoi{10.1086/309303}

\bibitem[{{Scherrer} {et~al.}(2012){Scherrer}, {Schou}, {Bush}, {Kosovichev},
  {Bogart}, {Hoeksema}, {Liu}, {Duvall}, {Zhao}, {Title}, {Schrijver},
  {Tarbell}, \& {Tomczyk}}]{hmischerrer}
{Scherrer}, P.~H., {Schou}, J., {Bush}, R.~I., {et~al.} 2012, \solphys, 275,
  207, \dodoi{10.1007/s11207-011-9834-2}

\bibitem[{{Schou} {et~al.}(2012){Schou}, {Scherrer}, {Bush}, {Wachter},
  {Couvidat}, {Rabello-Soares}, {Bogart}, {Hoeksema}, {Liu}, {Duvall}, {Akin},
  {Allard}, {Miles}, {Rairden}, {Shine}, {Tarbell}, {Title}, {Wolfson},
  {Elmore}, {Norton}, \& {Tomczyk}}]{hmischou}
{Schou}, J., {Scherrer}, P.~H., {Bush}, R.~I., {et~al.} 2012, \solphys, 275,
  229, \dodoi{10.1007/s11207-011-9842-2}

\bibitem[{{Semel} \& {Skumanich}(1998)}]{1998A&A...331..383S}
{Semel}, M., \& {Skumanich}, A. 1998, \aap, 331, 383

\bibitem[{{Severnyi}(1965)}]{1965SvA.....9..171S}
{Severnyi}, A.~B. 1965, \sovast, 9, 171

\bibitem[{{Toriumi} {et~al.}(2017){Toriumi}, {Schrijver}, {Harra}, {Hudson}, \&
  {Nagashima}}]{toriumi17}
{Toriumi}, S., {Schrijver}, C.~J., {Harra}, L.~K., {Hudson}, H., \&
  {Nagashima}, K. 2017, \apj, 834, 56, \dodoi{10.3847/1538-4357/834/1/56}

\bibitem[{{Toriumi} \& {Wang}(2019)}]{2019LRSP...16....3T}
{Toriumi}, S., \& {Wang}, H. 2019, Living Reviews in Solar Physics, 16, 3,
  \dodoi{10.1007/s41116-019-0019-7}

\bibitem[{{T{\"o}r{\"o}k} {et~al.}(2014){T{\"o}r{\"o}k}, {Leake}, {Titov},
  {Archontis}, {Miki{\'c}}, {Linton}, {Dalmasse}, {Aulanier}, \&
  {Kliem}}]{torok14}
{T{\"o}r{\"o}k}, T., {Leake}, J.~E., {Titov}, V.~S., {et~al.} 2014, \apjl, 782,
  L10, \dodoi{10.1088/2041-8205/782/1/L10}

\bibitem[{{Tsap} {et~al.}(2022){Tsap}, {Stepanov}, \&
  {Kopylova}}]{2022ApJ...939..114T}
{Tsap}, Y.~T., {Stepanov}, A.~V., \& {Kopylova}, Y.~G. 2022, \apj, 939, 114,
  \dodoi{10.3847/1538-4357/ac9833}

\bibitem[{{Tziotziou} {et~al.}(2013){Tziotziou}, {Georgoulis}, \&
  {Liu}}]{tziotziou13}
{Tziotziou}, K., {Georgoulis}, M.~K., \& {Liu}, Y. 2013, \apj, 772, 115,
  \dodoi{10.1088/0004-637X/772/2/115}

\bibitem[{{van Driel-Gesztelyi} \& {Green}(2015)}]{2015LRSP...12....1V}
{van Driel-Gesztelyi}, L., \& {Green}, L.~M. 2015, Living Reviews in Solar
  Physics, 12, 1, \dodoi{10.1007/lrsp-2015-1}

\bibitem[{{Vemareddy}(2019)}]{2019MNRAS.486.4936V}
{Vemareddy}, P. 2019, \mnras, 486, 4936, \dodoi{10.1093/mnras/stz1020}

\bibitem[{{Vemareddy} {et~al.}(2015){Vemareddy}, {Venkatakrishnan}, \&
  {Karthikreddy}}]{2015RAA....15.1547V}
{Vemareddy}, P., {Venkatakrishnan}, P., \& {Karthikreddy}, S. 2015, Research in
  Astronomy and Astrophysics, 15, 1547, \dodoi{10.1088/1674-4527/15/9/011}

\bibitem[{{Venkatakrishnan} \& {Tiwari}(2009)}]{2009ApJ...706L.114V}
{Venkatakrishnan}, P., \& {Tiwari}, S.~K. 2009, \apjl, 706, L114,
  \dodoi{10.1088/0004-637X/706/1/L114}

\bibitem[{{Verma}(2018)}]{2018A&A...612A.101V}
{Verma}, M. 2018, \aap, 612, A101, \dodoi{10.1051/0004-6361/201732214}

\bibitem[{{{\v{S}}vanda} {et~al.}(2021){{\v{S}}vanda}, {Sobotka},
  {Mravcov{\'a}}, \& {V{\'y}bo{\v{s}}{\v{t}}okov{\'a}}}]{2021A&A...647A.146S}
{{\v{S}}vanda}, M., {Sobotka}, M., {Mravcov{\'a}}, L., \&
  {V{\'y}bo{\v{s}}{\v{t}}okov{\'a}}, T. 2021, \aap, 647, A146,
  \dodoi{10.1051/0004-6361/202040127}

\bibitem[{{Wang} {et~al.}(2023){Wang}, {Qiu}, {Liu}, {Zhu}, {Yang}, {Hu}, \&
  {Wang}}]{2023ApJ...943...80W}
{Wang}, W., {Qiu}, J., {Liu}, R., {et~al.} 2023, \apj, 943, 80,
  \dodoi{10.3847/1538-4357/aca6e1}

\bibitem[{{Wheatland}(2000)}]{2000ApJ...532..616W}
{Wheatland}, M.~S. 2000, \apj, 532, 616, \dodoi{10.1086/308577}

\bibitem[{{Wilkinson} {et~al.}(1992){Wilkinson}, {Emslie}, \&
  {Gary}}]{wilkinson92}
{Wilkinson}, L.~K., {Emslie}, A.~G., \& {Gary}, G.~A. 1992, \apjl, 392, L39,
  \dodoi{10.1086/186420}

\bibitem[{{Zwaan}(1985)}]{1985SoPh..100..397Z}
{Zwaan}, C. 1985, \solphys, 100, 397, \dodoi{10.1007/BF00158438}

\bibitem[{{Zwaan}(1987)}]{1987ARA&A..25...83Z}
---. 1987, \araa, 25, 83, \dodoi{10.1146/annurev.aa.25.090187.000503}

\end{thebibliography}
\bibliographystyle{aasjournal}

\end{document}